\documentclass[letterpaper,onecolumn,10pt,accepted=2021-02-04]{quantumarticle}
\pdfoutput=1
\usepackage[numbers,sort&compress]{natbib}
\usepackage{amsmath, amssymb, graphicx, bm}

\newcommand{\br}{\bm{r}}

\newcommand{\hbr}{\hat{\bm{r}}}

\newcommand{\W}{\mathcal{W}}
\newcommand{\T}{\mathcal{T}}

\newcommand{\uu}{\mathcal{U}}
\renewcommand{\S}{\mathcal{S}}
\newcommand{\beq}{\begin{equation}}
\newcommand{\eeq}{\end{equation}}
\newcommand{\y}{y}

\begin{document}
\title[]{Exact and approximate continuous-variable \newline gate decompositions}

\author{Timjan Kalajdzievski}
\affiliation{Xanadu, Toronto, ON, M5G 2C8, Canada}
\author{Nicol\'as Quesada}
\affiliation{Xanadu, Toronto, ON, M5G 2C8, Canada}
\maketitle
\begin{abstract}
We gather and examine in detail gate decomposition techniques for continuous-variable quantum computers and also introduce some new techniques which expand on these methods. Both exact and approximate decomposition methods are studied and gate counts are compared for some common operations. While each having distinct advantages, we find that exact decompositions have lower gate counts whereas approximate techniques can cover decompositions for all continuous-variable operations but require significant circuit depth for a modest precision.  
\end{abstract}

\section{Introduction}
For a quantum computer to be able to execute an algorithm, often defined by a product of high-level unitary transformations \cite{ChrisAlgo2017, kalajdzievski2018continuous, CVHHL, Loock2013, SolovayReview, amy2013meet}, the operations of the algorithm must be translated into the small set of logical gates directly built into the hardware. These built-in operations are the quantum computing analog to the logical gates which are hardwired into the transistors of a classical computer. In order to program a quantum computer to execute an algorithm, a combination of the built-in elementary gates must be found that can reproduce the operation of the algorithm. Certain sets of gates exist such that any arbitrary unitary operation can be expressed as a finite product of gates from the set, to any desired precision \cite{lloyd1995almost, divincenzo1995two, barenco1995elementary, SethUniversal, SethCV, bravyi2005universal}. Due to their ability to express any given transformation, these are referred to as universal gates sets. Thus, in order to program a quantum computer, a method is required to decompose any given unitary in terms of elements of the universal gate sets. Ideally, a decomposition method will allow for the implementation of an algorithm with high precision and as few gates as possible. Gate decomposition methods and techniques have been broadly studied, especially in the discrete variable (qubit) domain \cite{lloyd1995almost, SethUniversal, CommutatorApprox, ExactDecomp, kitaev1997quantum}. 

The qubit model of quantum computing uses two-level quantum systems that act as the quantum analog of bits on a classical computer. The logical gates acting on qubits can be represented by $2 \times 2$ unitary matrices, and the process of translating general operations into the logical gates of this model, or gate decomposition, has been well studied. For example, the Solovay-Kitaev theorem \cite{kitaev1997quantum, SolovayReview}, stated informally tells us that any single-qubit operation can be approximated to high precision using a sequence of only a few logical gates. Other results have also been discovered which strengthen decomposition techniques in qubit systems by allowing for less logical gates to be needed, or the resulting decomposition to be of a greater precision. These techniques are used for single-qubit operations \cite{kliuchnikov2013asymptotically, kliuchnikov2014asymptotically, kliuchnikov2015framework, bocharov2015efficient} as well as general multi-qubit operations \cite{amy2013meet}.

In this manuscript we focus on gate decomposition techniques for continuous-variable (CV) quantum computers, which have not been as extensively studied. In the CV model of quantum computing, registers are infinite-dimensional quantum systems called qumodes (the infinite-dimensional generalization of qubits) and physically correspond to quantum harmonic oscillators. The logic gates in this model are unitary operators which act on the infinite-dimensional Hilbert space \cite{SethCV, Menicucci2006, Gu2009, ChrisOverview, ChrisAlgo2017, kalajdzievski2018continuous, CVHHL, Loock2013}. CV quantum computing using quadrature operators to form logic gates was first proposed by Lloyd and Braunstein \cite{SethCV}. The logic gates in that case were formed using Hamiltonians which were polynomial in the quadrature operators. These gates allow one to express analog computation using quantum-mechanical operations, and moreover, by constructing a universal set one can construct any gate generated by a Hamiltonian that is a polynomial of finite order in the quadrature operations, like the ones typically found in the simulation of bosonic systems \cite{BHExtended, kalajdzievski2018continuous}. The problem of finding optimal ways to decompose an arbitrary gate into a product of gates from a given universal set is precisely the focus of this manuscript. The study of this problem also finds application when continuous degrees of freedom are used to encode error-correctable qubits, since the logical operations of the logical qubits need to be ultimately executed as continuous-variable gates \cite{myers2011coherent, ralph2003quantum, pantaleoni2020modular, gottesman2001encoding}.


We have structured our manuscript as follows: in Sec.~\ref{sec:background} we set up some basic notations. Then in  Sec.~\ref{Commutator} we examine a comprehensive method for decomposing arbitrary CV transformations using commutator identities introduced by Sefi and van Loock in Ref.~\cite{CommutatorApprox}. This method is approximate, meaning that the final decomposition only implements the desired operation up to a certain error. The error is decreased by repeating the sequence of gates in the decomposition \cite{Suzuki2005, CommutatorApprox, Wiebe2010}. Exact decompositions which do not rely on Trotterization (see next section) are also known but are not universally applicable \cite{ChrisAlgo2017, CommutatorApprox, CVHHL}. We first study exact decompositions of quadratic Hamiltonians in Sec.~\ref{gaussian-decomp} using their symplectic representation~\cite{braunstein2005squeezing,dutta1995real}. Then, in Sec.~\ref{higher1} we study exact decomposition of higher than quadratic polynomials of mutually commuting operators \cite{ExactDecomp, PhDThesis}. These methods are further extended and generalized in Sec.~\ref{higher2}.
In Sec. \ref{SqDis} we examine another approximate method which relies on using squeezing and displacement operations to assemble a single mode cubic-phase gate starting from a Kerr gate~\cite{MabuchiKerr}. Finally we provide a comparison of the methods as well as some discussion in Sec. \ref{discussion}.


%

\section{Background and notation}~\label{sec:background}

Registers in the CV model are quantum harmonic oscillators with corresponding creation and annihilation operators $\hat{a}^{\dagger}_j$ and $\hat{a}_j$, where the subscript refers to the mode they act upon. For a system with $M$ modes, these operators satisfy the bosonic commutation relations $[\hat{a}_j , \hat{a}^{\dagger}_k]=\delta_{jk}$, and $[\hat{a}_j, \hat{a}_k]=[\hat{a}^{\dagger}_j, \hat{a}^{\dagger}_k]=0$. An equivalent operator description of a bosonic system uses the quadrature operators $\hat{x}$ and $\hat{p}$, which can be represented in terms of the annihilation and creation operators as
\begin{align}
\hat{x}_{j} = \sqrt{\frac{\hbar}{2}}\left(\hat{a}^{\dagger}_j+ \hat{a}_j\right),\quad \hat{p}_{j} = i \sqrt{\frac{\hbar}{2}}\left(\hat{a}^{\dagger}_j - \hat{a}_j\right), \label{Expand}
\end{align} 
with commutators $[\hat{x}_j,\hat{x}_k]=[\hat{p}_j,\hat{p}_k]=0$ and  $[\hat{x}_j,\hat{p}_k] = i \hbar \delta_{j,k}$. 
For convenience we group the operators as follows,
\begin{align}
\bm{\hat x} &= (\hat x_1,\ldots,\hat x_M)^T, \quad \bm{\hat p} = (\hat p_1,\ldots,\hat p_M)^T, \\
\bm{\hat r} &= \begin{pmatrix}
\hat{\bm{x}} \\
\hat{\bm{p}}
\end{pmatrix}= (\hat x_1,\ldots,\hat x_M, \hat p_1,\ldots,\hat p_M)^T.
\end{align}
This notation allows us to write the commutation relations as $[\hat{r}_j,\hat{r}_k] = i \hbar \Omega_{j,k}$ where 
\begin{align}\label{eq:sympform}
\bm{\Omega} = \begin{pmatrix}
0 & \ \mathbb{I}_M \\
-\mathbb{I}_M & \ 0
\end{pmatrix}
\end{align}
is the so-called symplectic form, and $\mathbb{I}_M$ is the $M$-dimensional identity matrix.



As mentioned previously, a universal gate set is a collection of gates such that any arbitrary unitary operation can be expressed as a finite sequence of gates from the set, to any desired precision. The CV decomposition techniques examined in later sections focus on the universal set specified by the gates

\begin{subequations} \label{Uniset}
\begin{align}
    R_j(\theta) &= \exp\left(i \tfrac{\theta}{2 \hbar} \left[ \hat{x}_j^2+ \hat{p}_j^2\right] \right), \label{eq:Rgatedef} \\
    Z_j(t_1) & = \exp\left(i \frac{t_1}{\hbar} \hat{x}_j \right), \label{eq:Zgatedef} \\
    P_j(t_2) & = \exp\left(i \frac{t_2}{2\hbar} \hat{x}^{2}_j \right), \label{eq:Pgatedef} \\
    V_j(t_3) & = \exp\left(i \frac{t_3}{3\hbar} \hat{x}^{3}_j \right), \label{eq:Vgatedef} \\
    CZ(\tau) & = \exp\left(i \frac{\tau}{\hbar} \hat{x}_j \hat{x}_k \right), \label{eq:CZgatedef}
\end{align}
\end{subequations}
where $\theta$, $t_{1}$, $t_{2}$, $t_{3}$, and $\tau$ are real parameters. This particular universal set is often chosen for mathematical convenience. Gates generated by a Hamiltonian that is at most quadratic are referred to as Gaussian operations. In the universal set above the rotation operator (\ref{eq:Rgatedef}), pure-momentum displacement (\ref{eq:Zgatedef}), quadratic-phase (\ref{eq:Pgatedef}), and the multi-mode controlled-phase gate (\ref{eq:CZgatedef}) are all Gaussian. These operations are implemented experimentally with linear optics.

The cubic phase gate (\ref{eq:Vgatedef}), which is the only non-Gaussian operator in the universal set, provides the non-linearity needed to allow for decompositions of arbitrarily higher order. This gate can be realized either by using optical nonlinearities \cite{MabuchiKerr} or measurements \cite{gottesman2001encoding,yukawa2013generation,gu2009quantum,marshall2015repeat,sabapathy2018states,sabapathy2019production, Cubic2018}. For example, in \cite{gottesman2001encoding} a photon counting measurement is used to introduce the higher order non-linearity. In this case the full implementation involves a displaced two-mode squeezed state for which $R^\dagger \hat n R$ (photon counting in a rotated basis) is measured on one arm. The desired cubic operation is then collapsed onto the second unmeasured mode. In \cite{marshall2015repeat} repeat-until-success photon subtractions and Gaussian operations are used, and in \cite{Cubic2018} quadrature detection for feed-forward manipulation of parameters produces the necessary nonlinear interaction. The cubic phase gate can also be realized in the microwave domain using the Josephson nonlinearity \cite{hillmann2020universal}.

A rotation operation by $\theta = \frac{\pi}{2}$ is referred to as the Fourier transform gate \cite{SethQFT, ManasQFT}. This operation maps between the quadrature operators $\hat{x}$ and $\hat{p}$ and is a cheap gate to implement experimentally. 
\begin{align}\label{Fourier}
\mathcal{F}^{\dagger}\hat{x}\mathcal{F}=-\hat{p},\quad \mathcal{F}^{\dagger}\hat{p}\mathcal{F}=\hat{x}.
\end{align}

An equivalent universal set is one where the non-Gaussian gate $V_j(t_3)$ is replaced by the Kerr gate $\exp\left( i t_3 (\hat{a}^\dagger \hat{a})^2 \right)$. In general, the non-Gaussian gates in the set can be replaced with any other and the set will retain universality. The reason why the cubic phase gate is chosen in the universal set here, as opposed to a Kerr gate, is due to mathematical convenience in the exact decompositions. As well, at the moment it seems as though gates with a mixture of $\hat{x}$ and $\hat{p}$ in the same mode cause problems for these same methods. This is because the polynomials used to create the desired powers in the decompositions will no longer be able to be constructed with unitary conjugation. Finally, note that in Sec.~\ref{SqDis} we examine in detail a method derived in Ref.~\cite{MabuchiKerr} where a driven-Kerr Hamiltonian is used to synthesize a cubic phase gate by unitary conjugation with squeezing and displacement gates. 

One can also remove one of the Gaussian gates by showing that it can be decomposed using other elements in the set \cite{CommutatorApprox}. In both of these cases the choice is often made such that the gates in the set are not costly to implement experimentally. If the decomposition of one gate in the set includes multiple uses of another harder to implement gate, then it might be beneficial to still include it. For example, the quadratic phase gate can be decomposed in terms of other gates, but its decomposition includes multiple uses of the cubic phase gate. This means that it can be taken out of the universal set, but as the cubic phase gate is much more costly to implement, it is common to still include it. In general, Gaussian gates may be decomposed exactly in terms of squeezing, displacements and beamsplitters. We demonstrate how to achieve these decompositions in section \ref{gaussian-decomp}. 

 
Throughout this manuscript we express an arbitrary unitary operator as $U = e^{itH}$ where the Hermitian operator $H=\sum_{j=1}^N H_{j}$ is the generator of the gate. When decomposing arbitrary gates into a universal set, it is often necessary to approximate this sum of operators in the exponent as a product of exponential operators. More specifically, for $H = \hat{A} + \hat{B}$ where $\hat{A}$ and $\hat{B}$ are Hermitian operators, the Zassenhaus formula \cite{magnus1954exponential} states that
\beq \label{splitting}
e^{it(\hat{A}+\hat{B})} = e^{it\hat{A}}e^{it\hat{B}}e^{\frac{t^2}{2}[\hat{A},\hat{B}]}e^{\frac{-it^3}{6}\left(2[\hat{B},[\hat{A},\hat{B}]]+[\hat{A},[\hat{A},\hat{B}]]\right)}\cdots
\eeq
When $[\hat{A},\hat{B}]=0$ the product ends immediately after the first two operations. In general, however, it is possible that this product never terminates, resulting in a decomposition that is no longer finite. In this case, the product can be truncated at a designated stage in the expansion and the remaining commutators can be neglected. This strategy is referred to as a Trotter-Suzuki approximation \cite{trotter1959product,Suzuki1976}, which in general can be written as
\beq\label{Eq: trotter-suzuki}
e^{itH}= \left(\prod_{j=1}^Ne^{i\frac{t}{K}H_j}\right)^K+O(t^2/K),
\eeq 
where $H = \sum_{j=1}^N H_{j}$. This approximation requires $K=O(1/\varepsilon)$ gates to achieve precision $\varepsilon$ for fixed $t$. Note that the subscript $j$ on $H_j$ is only an index and does not refer to a mode, as each $H_j$ may contain any number of modes. In general, the Trotter-Suzuki approximation can be useful in combination with a variety of other decomposition techniques and approximations as we show in the next section.

\section{Approximate decompositions using commutators} \label{Commutator}

In this section we examine an approximate decomposition technique detailed in Ref.~\cite{CommutatorApprox}. This method expresses sums and products of the quadrature operators in terms of commutators and then approximates the exponentials of these commutators as repeated products of their arguments. More specifically, given a general unitary $U = e^{it\hat{H}}$ where $\hat{H}$ is a sum of products of quadrature operators, each term in the sum can be represented in terms of commutators with the following identities:
\begin{align}\label{comm1}
\hat{x}^m &= -\frac{2}{3(m-1) (2 \hbar)^2}[\hat{x}^{m-1},[\hat{x}^3 , \hat{p}^2]], \\
\label{comm2} \hat{x}^m\hat{p}^n + \hat{p}^n\hat{x}^m &= -\frac{4i}{(n+1)(m+1)(2 \hbar)}[\hat{x}^{m+1},\hat{p}^{n+1}] - \frac{1}{(n+1)(2\hbar)^2}\sum_{k=1}^{n-1}[\hat{p}^{n-k},[\hat{x}^m,\hat{p}^k]],
\end{align}
and for two-mode terms
\begin{align}\label{comm3}
\hat{p}_{1}^m\hat{p}_{2}^n = -\frac{1}{(m+1)(n+1)(2 \hbar)^2}[\hat{p}_{2}^{n+1},[\hat{p}_{1}^{m+1} , \hat{x}_1\hat{x}_2]].
\end{align}
Once the exponent of the general unitary operator is written in terms of a sum of commutators, it can be separated with the Trotter-Suzuki technique as in Eq.~(\ref{Eq: trotter-suzuki}), and then expanded in the following way: given two Hermitian operators $\hat{A}$ and $\hat{B}$, it holds that \cite{childs2017toward}
\begin{align}\label{Eq: comm_approx}
e^{t^2[\hat{A},\hat{B}]}&=  \left(e^{i\frac{t}{K}\hat{B}}e^{i\frac{t}{K}\hat{A}}e^{-i\frac{t}{K}\hat{B}}e^{-i\frac{t}{K}\hat{A}}\right)^{K^2}+O(t^4/K).
\end{align}
Note that for fixed $t$, $K=O(1/\varepsilon)$ gates are required to achieve an error of $\varepsilon$ in this approximation, but the resulting circuit will have a depth of $O(1/\varepsilon^2)$. It is possible then that very large circuits may be required to achieve a desired precision. A similar expression can also be used in the case of a nested commutator in the form $e^{t^3[\hat{B},[\hat{B},\hat{A}]]}$, given by
\begin{align}\label{Eq: comm_approx_nest}
e^{it^3[\hat{B},[\hat{B},\hat{A}]]}&= \left(e^{i\frac{t}{K}\hat{B}}e^{\frac{t^2}{K^2}[\hat{B},\hat{A}]}e^{-i\frac{t}{K}\hat{B}}e^{-\frac{t^2}{K^2}[\hat{B},\hat{A}]}\right)^{K^2}+O(t^4/K).
\end{align}
To illustrate the use of this decomposition technique, consider the decomposition of the unitary $e^{it\left(\hat{x}^2\hat{p} + \hat{p}\hat{x}^2\right)}$. First, using the equality from Eq.~(\ref{comm2}) we have that
\beq
e^{it\left(\hat{x}^2\hat{p} + \hat{p}\hat{x}^2\right)} = e^{\frac{t}{3 \hbar}[\hat{x}^3,\hat{p}^2]}.
\eeq
Next, using Eq.~(\ref{Eq: comm_approx}) with $\hat{A}=\hat{x}^3$ and $\hat{B}=\hat{p}^2$ leads to
\begin{align} \label{commexample}
e^{\frac{t}{3 \hbar}[\hat{x}^3,\hat{p}^2]} \approx & \left(e^{i\frac{\sqrt{\frac{t}{3\hbar}}}{K}\hat{p}^2}e^{i\frac{\sqrt{\frac{t}{3\hbar}}}{K}\hat{x}^3}e^{-i\frac{\sqrt{\frac{t}{3\hbar}}}{K}\hat{p}^2}e^{-i\frac{\sqrt{\frac{t}{3\hbar}}}{K}\hat{x}^3}\right)^{K^2} + O\left[\left(\frac{t}{3 \hbar}\right)^2/K\right],
\end{align}
where each of the resulting gates on the right-hand side are part of the universal set in Eq.~(\ref{Uniset}) up to Fourier transforms. In order to obtain a precision of $O(1/K)$, this product of gates must be repeated $O(K^2)$ times. For example, if $t=1$ and the desired precision of the decomposition is $10^{-3}$, then the product of gates on the right-hand side needs to be repeated approximately $10^5$ times. \\

Another operator we can decompose to demonstrate this technique is the Kerr operation. We also examine the decomposition of this operation with a different method in section \ref{SqDis}. The Kerr operation in terms of quadrature operators can be written as $e^{it\left(\hat{x}^{4} + \hat{x}^{2}\hat{p}^{2} +\hat{p}^{2}\hat{x}^{2} + \hat{p}^{4} - \hbar\hat{x}^{2} - \hbar\hat{p}^{2} \right)}$. After splitting the terms in the exponent using Trotter-Suzuki as in Eq.~(\ref{Eq: trotter-suzuki}), and neglecting Fourier transforms, we are left to decompose operators which look like $e^{it\hat{x}^{4}}$ and $e^{it(\hat{x}^{2}\hat{p}^{2} +\hat{p}^{2}\hat{x}^{2})}$. These operators can be written in terms of commutators with Eq.~(\ref{comm1}) and ({\ref{comm2}) respectively to get
\begin{align}
e^{it\hat{x}^{4}} = e^{-\frac{it}{18\hbar^{2}}[\hat{x}^{3},[\hat{x}^{3},\hat{p}^{2}]]},
\end{align}
and
\begin{align}
e^{it(\hat{x}^{2}\hat{p}^{2} +\hat{p}^{2}\hat{x}^{2})} = e^{\frac{2t}{9\hbar}[\hat{x}^{3},\hat{p}^{3}]}.
\end{align}
Expanding with Eq.~(\ref{Eq: comm_approx}) and (\ref{Eq: comm_approx_nest}), we get
\begin{align} \label{commexexp}
&e^{-\frac{it}{18\hbar^{2}}[\hat{x}^{3},[\hat{x}^{3},\hat{p}^{2}]]} \\
&\ \ \approx 
 \left(e^{-i\frac{\left(\frac{t}{18\hbar^2}\right)^{1/3}}{K}\hat{x}^3}e^{\frac{\left(\frac{t}{18\hbar^2}\right)^{2/3}}{K^2}[\hat{x}^3,\hat{p}^2]}e^{i\frac{\left(\frac{t}{18\hbar^2}\right)^{1/3}}{K}\hat{x}^3}e^{-\frac{\left(\frac{t}{18\hbar^2}\right)^{2/3}}{K^2}[\hat{x}^3,\hat{p}^2]}\right)^{K^2} + O\left[\left(\frac{t}{18 \hbar^2}\right)^{4/3}/K\right], \nonumber 
\end{align}
and
\begin{align}
e^{\frac{2t}{9\hbar}[\hat{x}^{3},\hat{p}^{3}]} \approx & \left(e^{i\frac{\sqrt{\frac{2t}{9\hbar}}}{K}\hat{p}^3}e^{i\frac{\sqrt{\frac{2t}{9\hbar}}}{K}\hat{x}^3}e^{-i\frac{\sqrt{\frac{2t}{9\hbar}}}{K}\hat{p}^3}e^{-i\frac{\sqrt{\frac{2t}{9\hbar}}}{K}\hat{x}^3}\right)^{K^2} + O\left[\left(\frac{2t}{9 \hbar}\right)^2/K\right],
\end{align}
where the second and fourth operators on the right-hand side of Eq.~(\ref{commexexp}) can be expanded further, similar to the expansion in Eq.~(\ref{commexample}). Once again the product of gates on the right-hand sides of each equation must be repeated $O(K^2)$ times to obtain a precision of $O(1/K)$. In this case, because the original operator needs to be expanded first with a Trotter-Suzuki approximation, the final sequence of gates needs to again be repeated to increase the precision as in Eq.~(\ref{Eq: trotter-suzuki}). For derivations of some of these identities, as well as discussion on the necessity of good approximations, see the supplementary materials section of \cite{CommutatorApprox}

To potentially avoid the cost of repeating gates for greater precision, exact decomposition methods can be used as we show in the next section.

\section{Exact decompositions} \label{exactdecomps}

\subsection{Gaussian operations} \label{gaussian-decomp}
In this section we provide an exact decomposition for an arbitrary Gaussian operator in terms of displacement, interferometer and squeezing gates \cite{barnett2002methods}. Before developing the decompositions we provide definitions for these unitary operations.
We start with the single mode displacement operator,
\begin{align}
D(\alpha) = \exp( \alpha \hat{a}^\dagger - \alpha^* \hat{a}).
\end{align}
which takes as argument the complex number $\alpha$. The annihilation and quadrature operators transform as 
\begin{align} \label{disptransform}
D^\dagger(\alpha) \hat{a}D(\alpha)  = \hat{a} +\alpha, \quad  D^\dagger(\alpha) \hat{x} D(\alpha) = \hat{x} + \sqrt{2 \hbar} \Re(\alpha ), \quad D^\dagger(\alpha) \hat{p} D(\alpha) = \hat{p} + \sqrt{2 \hbar} \Im(\alpha ),
\end{align}
where $\alpha = \Re(\alpha ) + i \Im(\alpha )$. Note in particular that the pure-momentum displacement in Eq.~\eqref{eq:Zgatedef} is simply a displacement operator by a purely imaginary amount, $Z(t_1) = D(i t_1/\sqrt{2 \hbar })$. We also define a multimode displacement operator as follows
\begin{align}
\W(\bm{r}) = \bigotimes_{j=1}^M D_j\left( \tfrac{1}{\sqrt{2 \hbar}}(x_j+i p_j) \right), \text{ where }
\bm{r} = (x_1,\ldots, x_M,p_1\ldots,p_M)^T.
\end{align}
By inserting the prefactor  $\tfrac{1}{\sqrt{2 \hbar}}$ in the argument of the single-mode displacement operators we make the following relation $\hbar$-independent
\begin{align}
\W^\dagger(\bm{r}) \bm{\hat r} \W (\bm{r}) = \bm{\hat r}+\bm{r}.
\end{align}
We continue with the single-mode squeezing operator defined as
\begin{align}\label{sq}
\T(s) = \exp\left[ \tfrac{s}{2} \left( \hat{a}^{\dagger 2} -\hat{a}^2\right) \right],
\end{align}
which transforms the annihilation and quadrature operators as follows
\begin{align}
\T^\dagger(s) \hat{a} \T(s) = \cosh(s) \hat{a} + \sinh(s) \hat{a}^\dagger, \quad \T^\dagger(s) \hat{x} \T(s) = e^{s} \hat{x}, \quad \T^\dagger(s) \hat{p} \T(s) = e^{-s} \hat{p}.
\end{align}
Finally, a multi-mode interferometer operator $\uu$, parametrized by a unitary matrix $\bm{U}$ acts as follows
\begin{align}
\uu(\bm{U})^\dagger a_i \uu(\bm{U}) &= \sum_{i=1}^M U_{il} a_l,\\
\uu(\bm{U})^\dagger \hbr \  \uu(\bm{U}) &= \bm{O}_{\bm{U}} \hbr, \quad \bm{O}_{\bm{U}} \equiv \left(\begin{array}{c|c}
\Re\left(\bm{U} \right) & -\Im\left(\bm{U}\right) \\
\hline
\Im\left(\bm{U}\right) & \Re\left(\bm{U}\right)\label{inter}
\end{array}\right).
\end{align}
Note that the interferometer operators respect the group structure of the unitary group. 
This last property makes it easy to decompose an arbitrary $M$-mode interferometer $\uu(\bm{U})$ into a product of interferometers acting on at most two modes by simply decomposing the associated matrix $\bm{U}$ into a product of unitary matrices where each unitary matrix couples at most two modes
\cite{reck1994experimental,clements2016optimal,de2018simple,su2019hybrid}.

With this notation we are ready to tackle the decomposition of the most general operation generated by a quadratic Hamiltonian. We essentially follow the elegant arguments exposed by Serafini~\cite{CVQuantum2017}
and write this operator as
\begin{align}
\mathcal{Q}(\bm{H}, \bar{\bm{r}}) = \exp\left[ \tfrac{i}{\hbar} \left( \tfrac{1}{2} \hat{\bm{r}}^T \bm{H} \hat{\bm{r}} + \hat{\bm{r}}^T \bar{\bm{r}}\right)\right],
\end{align}
where we assume $\bm{H} \in \mathbb{R}^{2M \times 2M}$ is a symmetric matrix and  $\bar{\bm{r}} \in \mathbb{R}^{2M}$. To obtain a decomposition we first eliminate the linear part in the exponential by inserting resolutions of the identity in terms of $\W$ as follows
\begin{align}
\mathcal{Q}(\bm{H}, \bar{\bm{r}}) &= \W(\br) \W^\dagger(\br) \mathcal{Q}(\bm{H}, \bar{\bm{r}}) \W(\br) \W^\dagger(\br)  =  \W(\br) \exp\left[ \tfrac{i}{\hbar} \left( \tfrac{1}{2} (\hat{\bm{r}} + \br)^T \bm{H} (\hat{\bm{r}} + \br) + (\hat{\bm{r}}+\br)^T \bar{\bm{r}}\right)\right]\W^\dagger (\br) \nonumber \\
&=e^{\frac{i}{2 \hbar} \br^T \bm{H} \br +\frac{i}{\hbar } \br^T \bar{\br}} \W( \br ) \exp\left[ \tfrac{i}{\hbar} \left( \tfrac{1}{2} \hat{\bm{r}}^T \bm{H} \hat{\bm{r}} +\hbr^T\{\bar{\br} +  \bm{H} \br \} \right) \right]  \W^\dagger(\br).
\end{align}
Note that the first term in the right-hand side of the last equation is an inconsequential global phase and that if we pick as displacement $\br = -\bm{H}^{-1} \bar{\br}$ then the term inside the curly braces in the last equation vanishes and we are only left with an operator with a purely quadratic generator,
\begin{align}
\S(\bm{H}) = \exp\left[  \tfrac{i}{2 \hbar } \hat{\bm{r}}^T \bm{H} \hat{\bm{r}}  \right].
\end{align}
To decompose this gate we study its action on the canonical operators to find (cf. Sec. 3.2.2 of Ref.~\cite{CVQuantum2017})
\begin{align}
 \S(\bm{H})^\dagger \hbr \S(\bm{H})   &= \bm{S} \hbr, \\
\bm{S} &= e^{- \bm{\Omega} \bm{H}}.
\end{align}
Note that $\mathcal{S}(\bm{H})$ is an operator acting on the Hilbert space of $M$ modes while $S= e^{- \bm{\Omega} \bm{H}}$ is $2M \times 2M$ matrix. It is not hard to show that the latter is a member of the real symplectic group \cite{dutta1995real}, i.e., that it satisfies
\begin{align}
\bm{S}\bm{\Omega} \bm{S}^{T} = \bm{\Omega},
\end{align}
where $\bm{\Omega}$ is the symplectic form in Eq.~\eqref{eq:sympform}. 
To continue with the decomposition of the operator $\hat{S}(\bm{H})$ we will now turn our attention to the singular-value decomposition of the matrix $\bm{S}$,
\begin{align}
\bm{S} = \bm{O}^{(1)} \bm{Z} \bm{O}^{(2)}.
\end{align}
Since $\bm{S}$ is symplectic, one can construct the orthogonal matrices $\bm{O}^{(1)}, \bm{O}^{(2)}$ to also be symplectic and satisfy (cf. Appendix B of Ref.~\cite{CVQuantum2017})
\begin{align}\label{Obm}
\bm{O}^{(i)} = \left(\begin{array}{c|c}
\Re\left(\bm{U}^{(i)} \right) & -\Im\left(\bm{U}^{(i)}\right) \\
\hline
\Im\left(\bm{U}^{(i)}\right) & \Re\left(\bm{U}^{(i)}\right)
\end{array}\right),
\end{align}
where $\bm{U}^{(i)}$ is a complex unitary matrix of size $M \times M$. This form is precisely the same form appearing in Eq.~\eqref{inter}. Invoking the membership of $\bm{S}$ in the symplectic group one can also show that the singular values of $\bm{S}$ come in pairs of the form $(z,1/z)$ and thus one can write the diagonal symplectic matrix $\bm{Z}$ as
\begin{align}\label{Zbm}
\bm{Z} = \text{diag}(z_1,\ldots,z_M,1/z_1,\ldots,1/z_M).
\end{align}
By comparing Eq.~\eqref{Obm} and Eq.~\eqref{Zbm} with Eq.~\eqref{inter} and Eq.~\eqref{sq} respectively one identifies~\cite{braunstein2005squeezing}
\begin{align}
\S( \bm{H}) = \uu(\bm{U}^{(1)}) \left[  \bigotimes_{j=1}^M \T_j(\log z_j)  \right]\uu(\bm{U}^{(2)}),
\end{align}
which concludes the decomposition, since one can verify that the operators on either of the equations above transform in exactly the same way the quadratures $\bm{\hat r}$.

As a first example consider the decomposition of the single-mode ($M=1$) CV-phase gate~\cite{fiuravsek2003unitary}
\begin{align}
P(s) = \exp\left( i\frac{s}{2 \hbar} \hat{x}^2 \right) = \exp\left( \frac{i}{2 \hbar } \hbr 
\begin{pmatrix}
s & 0 \\
0 & 0
\end{pmatrix}
\hbr^T \right),
\end{align} where without loss of generality we take $s\geq 0$, since ${P}(-s) = {P}^\dagger(s)$. 
From this definition we easily find 
\begin{align}
e^{- \bm{\Omega} \bm{H}} &= \begin{pmatrix}
1 & 0 \\
s & 1
\end{pmatrix}=
\begin{pmatrix}
\cos \theta & -\sin \theta \\
\sin \theta & \cos \theta 
\end{pmatrix}
\begin{pmatrix}
e^r & 0 \\
0 & e^{-r} 
\end{pmatrix}
\begin{pmatrix}
\sin \theta & \cos \theta \\
-\cos \theta & \sin \theta
\end{pmatrix}, \\
\frac{s}{2} &=  \sinh r, \quad \cos \theta = \frac{1}{\sqrt{1+e^{2r}}}, \quad \sin \theta = \frac{e^{r}}{\sqrt{1+e^{2r}}}, \label{BS1}
\end{align} 
and then we can write
\begin{align}
P(s) = R(\theta) \T(r) {R}(\theta-\pi/2),
\end{align}
where we used the fact that a single-mode interferometer $\uu(e^{i \theta})$  is identical to a rotation gate  ${R}(\theta)$.

As a second example let us consider the two-mode controlled-phase gate,
\begin{align}
&CZ(s) = \exp\left( i \tfrac{s}{\hbar} \hat{x}_1 \hat{x}_2 \right) = \exp\left[ \frac{i}{2\hbar } \hbr 
\left( \begin{smallmatrix}
0 & s & 0 & 0 \\
s & 0 & 0 & 0 \\
0 & 0 & 0 & 0 \\
0 & 0 & 0 & 0
\end{smallmatrix} \right)
\hbr^T \right], \\
&e^{- \bm{\Omega} \bm{H}} = \left(
\begin{array}{cccc}
 1 & 0 & 0 & 0 \\
 0 & 1 & 0 & 0 \\
 0 & s & 1 & 0 \\
 s & 0 & 0 & 1 \\
\end{array}
\right) \\
&= \left(
\begin{array}{cccc}
 \cos \theta  & 0 & 0 & -\sin \theta  \\
 0 & \cos \theta  & -\sin \theta  & 0 \\
 0 & \sin \theta  & \cos \theta  & 0 \\
 \sin \theta  & 0 & 0 & \cos \theta  \\
\end{array}
\right)
\left(
\begin{array}{cccc}
e^r & 0 & 0 & 0 \\
0 & e^r & 0 & 0 \\
0 & 0 & e^{-r} & 0 \\
0 & 0 & 0 & e^{-r} \\
\end{array}
\right)
\left(
\begin{array}{cccc}
 \sin \theta  & 0 & 0 & \cos \theta  \\
 0 & \sin \theta  & \cos \theta  & 0 \\
 0 & -\cos \theta  & \sin \theta  & 0 \\
 -\cos \theta  & 0 & 0 & \sin \theta  \\
\end{array}
\right),\nonumber 
\end{align}
where $r$ and $\theta$ have the same functional dependence on the parameter $s$ as in Eq.~\eqref{BS1} and we assumed without loss of generality that $s\geq 0$.
Having the singular-value decomposition of the symplectic matrix we can easily write
\begin{align}
CZ(s) =BS(\theta)  \left[ \T_1(r) \otimes \T_2(r) \right]  BS(\theta-\pi/2),
\end{align}
where $BS(\theta)$ is the (symmetric) beamsplitter
\begin{align}
BS(\theta) = \uu \left( \left[  \begin{smallmatrix}
\cos \theta & i \sin \theta \\
i \sin \theta & \cos \theta 
\end{smallmatrix} \right] \right)
 = \exp\left( i \theta \left[\hat a_1^\dagger \hat{a}_2+ \hat a_2^\dagger \hat{a}_1 \right] \right).
\end{align}
\subsection{Non-Gaussian  Operators}\label{higher1}
In this section we study exact decomposition techniques for non-Gaussian operators. The techniques apply broadly to gates that have generators made of monomials that mutually commute. 
These techniques were first introduced by one of the authors in Ref. \cite{ExactDecomp} and provides a recipe for exact decompositions of multi-mode gates $e^{itH}$, where the operator $H$ is of the form
\begin{align}\label{Eq: class_of_H}
H= \left(\prod_{j=1}^{N-1} \hat{x}_{j}\right)\hat{x}_N^n,
\end{align}
or 
\begin{align}\label{Eq: class_of_H2}
H=  \hat{x}_{1}^{n_1}\hat{x}_{2}^{n_2},
\end{align}
for $n$, $n_1$, and $n_2$ positive integers. As well as single-mode gates $e^{itH}$ with
\beq
H=\hat{x}^{N}.
\eeq
These gates can also be extended to include momentum quadrature operators $\hat{p}_j$ by Fourier transforms acting on individual modes. Note that the method works for any product where at most one operator has an exponent $n>1$, so the label of the modes in Eq.~\eqref{Eq: class_of_H} is arbitrary. It is also required that $N$ is divisible by either 2 or 3, and in the multi-mode case, the product $nN$ is divisible by 2 or 3, as well as $n_1$ and $n_2$ divisible by 2. The intuition behind these restrictions is that the method relies on factoring the overall power $N$ into two lower powers and being able to decompose operators with exponents to those two lower powers. Because of the recursive nature of some of the single mode gates which are decomposed with this method, only decompositions of certain powers are known \cite{PhDThesis}. More specifically, all single mode gates with power divisible by 2 or 3 are known, and as a result, the overall powers of any decomposition must also be divisible by 2 or 3. Although more restricted than the approximate commutator expansion method, the set of gates for which these exact decompositions can be obtained encompasses a large class of operators arising in several CV quantum algorithms and simulations of bosonic systems \cite{sparrow2018simulating, CVHHL, kalajdzievski2018continuous, BHExtended, CommutatorApprox, ChrisAlgo2017, MonteCarlo}. \\ 

The method relies on using the following three identities: \\
(i) unitary conjugation
\begin{equation}\label{Eq:Unit_conj}
U^{\dagger}e^{it H} U= e^{it U^{\dagger}HU},
\end{equation}
(ii) a lemma to the Baker-Campbell-Hausdorff (BCH) formula
\begin{align}\label{Eq:BCH}
e^{\hat{A}} \hat{B} e^{-\hat{A}} = \hat{B} + [\hat{A},\hat{B}] +& \frac{1}{2!}[\hat{A},[\hat{A},\hat{B}]] + \ldots,
\end{align}
and (iii), the identity
\begin{align} \label{Unbalanced}
e^{i3\alpha^2 t \hat{p}_{k}\hat{x}_{j}^2 } = \: &e^{i\frac{\alpha}{\hbar} \hat{x}_{j} \hat{x}_{k}} e^{i\frac{t}{\hbar}\hat{p}_{k}^3}e^{-i\alpha \hat{x}_{j} \hat{x}_{k}}e^{-i\frac{t}{\hbar}\hat{p}_{k}^3}e^{-i\frac{\alpha}{\hbar} \hat{x}_{j} \hat{x}_{k}} e^{i\frac{t}{\hbar}\hat{p}_{k}^3}e^{i\alpha \hat{x}_{j} \hat{x}_{k}}e^{-i\frac{t}{\hbar}\hat{p}_{k}^3} e^{i\frac{3 \alpha^3 t}{4 \hbar}\hat{x}_{j}^3},
\end{align}
with $\alpha$ and $t$ real parameters. The identity Eq.~(\ref{Unbalanced}) comes from repeated usage of both (i) and (ii), the details of which are shown in Appendix A of Ref. \cite{PhDThesis}, as well as a few derivations of other similar identities in the appendix of that reference. A few examples can now be studied to provide intuition on how these identities can be used.\\

Assume an exact decomposition is needed for the unitary operator $e^{i\alpha\hat{x}_{j}\hat{x}_{k}\hat{x}_{l}}$. First, express the operator $\hat{x}_{j}\hat{x}_{k}\hat{x}_{l}$ as a linear combination of polynomials of degree three in the quadrature operators $\hat{x}_{j}$, $\hat{x}_{k}$, and $\hat{x}_{l}$. This is done using the identity
\begin{align}\label{ExThreeMode}
\hat{x}_{j}\hat{x}_{k}\hat{x}_{l}=&\tfrac{1}{6}[(\hat{x}_{j} + \hat{x}_{k} + \hat{x}_{l})^3-(\hat{x}_{j} + \hat{x}_{k})^3-(\hat{x}_{j} + \hat{x}_{l})^3-(\hat{x}_{k} + \hat{x}_{l})^3+\hat{x}_{j}^3+\hat{x}_{k}^3+\hat{x}_{l}^3],
\end{align}
which implies the identity
\begin{align} \label{DecompCube2}
e^{i\alpha\hat{x}_{j}\hat{x}_{k}\hat{x}_{l}} = \: &e^{\frac{i\alpha}{6}\left(\hat{x}_{j} + \hat{x}_{k} + \hat{x}_{l}\right)^3} e^{\frac{-i\alpha}{6}\left(\hat{x}_{j} + \hat{x}_{k}\right)^3}e^{\frac{-i\alpha}{6}\left(\hat{x}_{j} + \hat{x}_{l}\right)^3}e^{\frac{-i\alpha}{6}\left(\hat{x}_{k} + \hat{x}_{l}\right)^3}
e^{\frac{i\alpha}{6}\hat{x}^{3}_{j}} e^{\frac{i\alpha}{6}\hat{x}^{3}_{k}} e^{\frac{i\alpha}{6}\hat{x}^{3}_{l}},
\end{align}
since all the terms in the exponent commute. The right-hand side of this equation still includes operators which are not contained within the universal set. To decompose these, the identities in Eq.~(\ref{Eq:Unit_conj}) and Eq.~(\ref{Eq:BCH}) can be used with $U=e^{2i\hat{p}_{j}\hat{x}_{k}}$ as the unitary of conjugation, to arrive at the following decompositions
\begin{align}\label{ThreeModePoly1}
e^{i\alpha\left(\hat{x}_{j} + \hat{x}_{k}\right)^3} &= e^{i\hat{p}_{j}\hat{x}_{k}/\hbar} e^{i\alpha\hat{x}_{j}^3} e^{-i\hat{p}_{j}\hat{x}_{k}/\hbar},\\
\label{ThreeModePoly2} e^{i\alpha\left(\hat{x}_{j} + \hat{x}_{k}+ \hat{x}_{l}\right)^3} &= e^{i\hat{p}_{j}\hat{x}_{l}/\hbar} e^{i\alpha(\hat{x}_{j}+\hat{x}_k)^3} e^{-i\hat{p}_{j}\hat{x}_{l}/\hbar }.
\end{align}
To summarize this example, the exponent of the original operator $\hat{x}_{j}\hat{x}_{k}\hat{x}_{l}$ is expressed as a linear combination of polynomials of operators, and as such the full operator $e^{i\alpha\hat{x}_{j}\hat{x}_{k}\hat{x}_{l}}$ can be written in terms of a product of gates, each of which is decomposed exactly using two of the three identities mentioned above. \\

To illustrate the method for a higher order single-mode gate, assume the operator we wish to decompose is $e^{i\alpha\hat{x}_j^4}$. Following the strategy used for the previous example, express the operator $\hat{x}_j^4$ as a linear combination of degree-four polynomials. More specifically, the identity
\beq
\hat{x}_j^4 = (\hat{x}_{j}^2 + \hat{x}_{k})^2-\hat{x}_k^2-2\hat{x}_j^2\hat{x}_k,
\eeq
implies the relation
\beq \label{fourthorder}
e^{i\alpha\hat{x}_j^4}=e^{i\alpha(\hat{x}_{j}^2 + \hat{x}_{k})^2}e^{-i\alpha\hat{x}_k^2}e^{-2i\alpha\hat{x}_j^2\hat{x}_k}.
\eeq 
Here, the operator $e^{-i\alpha\hat{x}_k^2}$ is contained in the universal set, while Eq.~\eqref{Unbalanced} gives an exact decomposition for $e^{-i\alpha\hat{x}_j^2\hat{x}_k}$ up to a Fourier transform. Similar to the previous example, the remaining term can be decomposed using unitary conjugation:
\begin{equation}\label{Eq:19}
e^{i\alpha(\hat{x}_{j}^2 + \hat{x}_{k})^2}=e^{i\hat{p}_{k}\hat{x}_{j}^2 /\hbar} e^{i\alpha\hat{x}_{k}^2} e^{-i\hat{p}_{k}\hat{x}_{j}^2/\hbar},
\end{equation}
leading to a full decomposition for the target gate $e^{i \alpha \hat{x}_j^4}$. Note that an additional ancillary mode $k$ was required in this decomposition.

To extend the method to higher order single and multi-mode gates a more general form of Eq.~(\ref{Unbalanced}) is needed 
\begin{align}\label{twomode}
e^{2i\alpha^2 \hat{p}_{k}\hat{x}_{j}^{N}} = \: &e^{i\frac{\alpha}{\hbar} \hat{x}_{j}^{N-2} \hat{x}_{k}} e^{-i\alpha \hat{x}_{j}^2 \hat{p}_{k}^2} e^{-2i\frac{\alpha}{\hbar} \hat{x}_{j}^{N-2} \hat{x}_{k}} e^{i\alpha \hat{x}_{j}^2 \hat{p}_{k}^2} e^{i\frac{\alpha^3}{\hbar} \hat{x}_{j}^{2(N-1)}},
\end{align}
where decompositions of single mode gates of lower order in $N$ are required. Although, in this more general case the same strategy is used. Namely, the target gate is expressed in terms of a linear combination of polynomials and the resulting gates are decomposed using unitary conjugation or another lower order decomposition. The derivation of this identity can be found in the appendix of Ref. \cite{PhDThesis}.

The same procedure involving creating polynomials and increasing overall order can also be used to characterize decompositions for the group of operators with $H$ given by Eq.~(\ref{Eq: class_of_H2}). Although in this case the orders of each power must be increased individually through the build up of polynomials.

Note that this method can provide exact decompositions for a wide variety of useful operations \cite{sparrow2018simulating, CVHHL, kalajdzievski2018continuous, BHExtended, CommutatorApprox, ChrisAlgo2017, MonteCarlo}, but the total gate count of the decomposition scales exponentially as the overall power in each mode is increased.

\subsection{Generalizing exact decomposition methods}\label{higher2}
By allowing for additional constants to be added to the polynomials, the scaling of the method in the previous section can be significantly improved and the method extended to allow for the decomposition of a larger category of operators. This is shown in this section with the use of a general formula for representing products as sums from Ref. \cite{ProdSum}.

As was discussed in the previous section, one of the types of Hamiltonians which can be decomposed exactly is of the form $e^{it\hat{H}}$, where the operator $H$ is of the form
\begin{align}
H= \left(\prod_{j=1}^{n-1} \hat{x}_{j}\right)\hat{x}_n^{s_n}.
\end{align}
The method for decomposing these Hamiltonians worked for any product where at most one operator has an exponent $s_{n}>1$. It also required that $n$ is divisible by either 2 or 3, and that the product $ns_{n}$ must also be divisible by 2 or 3. 

A lemma from Ref. \cite{ProdSum} can be used to find decompositions which no longer have the first restriction that at most only one operator has an exponent $s_{n}>1$. In other words, the position quadrature operators in the product can be taken to any arbitrary powers, as long as the overall sum of the powers is divisible by 2 or 3.

The following equation is valid for real numbers \cite{ProdSum}, but because each of the $\hat{x}_{i}$s commute with each other it is also valid for a product of operators.
\begin{align} \label{GenProd} 
\hat{x}^{s_1}_1\hat{x}^{s_2}_2 \cdots \hat{x}^{s_n}_n = \frac{1}{s!}\sum_{v_{1}=0}^{s_{1}}\cdots \sum_{v_{n}=0}^{s_{n}} (-1)^{\sum_{i=1}^{n} v_{i}} \binom{s_1}{v_1} \cdots \binom{s_n}{v_n}\left(\sum_{i=1}^{n}h_i \hat{x}_{i}\right)^{s},
\end{align}
where $s=s_1 + s_2 + \cdots + s_n$ and $h_i = s_i /2 - v_i$. When at least one of the $s_i$ are odd repeated terms in the sum may be grouped together reducing the expression to
\begin{align} \label{OddProd}
\hat{x}^{s_1}_1\hat{x}^{s_2}_2 \cdots \hat{x}^{s_n}_n = \frac{2}{s!}\sum_{v_{1}=0}^{(s_{1}-1)/2}\sum_{v_{2}=0}^{s_{2}}\cdots \sum_{v_{n}=0}^{s_{n}} (-1)^{\sum_{i=1}^{n} v_{i}} \binom{s_1}{v_1} \cdots \binom{s_n}{v_n}\left(\sum_{i=1}^{n}h_i \hat{x}_{i}\right)^{s},
\end{align}
where without loss of generality it is assumed that $s_1$ is odd. Following the procedure in the last section, each term on the right-hand side may be created through unitary conjugation, where each operation in the unitary conjugation can be decomposed exactly. The challenge then is to be able to implement gates with the correct coefficients which depend on each $s_i$. We note that, to the best of our knowledge, the idea of using the identity in Eq.~\eqref{GenProd} to decompose CV gates has not been presented in the literature before.

As an example, expand the product $\hat{x}_1\hat{x}^{2}_2\hat{x}^{3}_3$ using Eq.~(\ref{OddProd}). Note that this product is not covered by the $H$ in Eq.~(\ref{Eq: class_of_H}), but has an overall power $s$ which is even.
\begin{align}
\hat{x}_1\hat{x}^{2}_2\hat{x}^{3}_3 & = \frac{1}{360}\sum_{v_{1}=0}^{0}\sum_{v_{2}=0}^{2}\sum_{v_{3}=0}^{3} (-1)^{\sum_{i=1}^{n} v_{i}} \binom{1}{v_1}\binom{2}{v_2}\binom{3}{v_3}\left(h_1 \hat{x}_{1} + h_2 \hat{x}_{2}+h_3 \hat{x}_{3} \right)^{6} \\ \nonumber
& = \frac{1}{360}\left(\hat{x}_{2} + \frac{1}{2}\hat{x}_{1} + \frac{3}{2}\hat{x}_{3}\right)^{6} - \frac{1}{120}\left(\hat{x}_{2} + \frac{1}{2}\hat{x}_{1} + \frac{1}{2}\hat{x}_{3}\right)^{6} + \frac{1}{120}\left(\hat{x}_{2} + \frac{1}{2}\hat{x}_{1} - \frac{1}{2}\hat{x}_{3}\right)^{6}  \\ \nonumber
& \; \; \; - \frac{1}{360}\left(\hat{x}_{2} + \frac{1}{2}\hat{x}_{1} - \frac{3}{2}\hat{x}_{3}\right)^{6} - \frac{1}{11520}\left(\hat{x}_{1} + 3\hat{x}_{3}\right)^{6} + \frac{1}{3840}\left(\hat{x}_{1} + \hat{x}_{3}\right)^{6} - \frac{1}{3840}\left(\hat{x}_{1}- \hat{x}_{3}\right)^{6}  \\ \nonumber 
& \; \; \; + \frac{1}{11520}\left(\hat{x}_{1} - 3\hat{x}_{3}\right)^{6} + \frac{1}{360}\left(\hat{x}_{2} - \frac{1}{2}\hat{x}_{1} - \frac{3}{2}\hat{x}_{3}\right)^{6} - \frac{1}{120}\left(\hat{x}_{2} - \frac{1}{2}\hat{x}_{1} - \frac{1}{2}\hat{x}_{3}\right)^{6} \\ \nonumber
& \; \; \; + \frac{1}{120}\left(\hat{x}_{2} - \frac{1}{2}\hat{x}_{1} + \frac{1}{2}\hat{x}_{3}\right)^{6} - \frac{1}{360}\left(\hat{x}_{2} - \frac{1}{2}\hat{x}_{1} + \frac{3}{2}\hat{x}_{3}\right)^{6}.
\end{align}
Each of the resulting terms can be made with unitary conjugation, where the central operator will have the single mode with unit coefficient in the brackets of each term. 


This relation along with the identities used in the last section (Eq.~(\ref{Eq:Unit_conj}), (\ref{Eq:BCH}), and (\ref{Unbalanced}) ) can also be used to find decompositions for operators that were already covered in the last section, in order to compare gate counts. For the operator where  $H = \hat{x}_{1}\hat{x}_{2}\hat{x}_{3}$ the total gate count using the previous method is 17 gates from the universal set, whereas when using the updated method described here it is 20 gates. For a higher order operation like $H = \hat{x}_{1}\hat{x}_{2}\hat{x}_{3}\hat{x}_{4}$ the total gate count from the universal set was 440 but with this method is 280. Other examples comparing the two exact methods are given in Table \ref{ExactVsComm}. In some cases it seems like the method in this section requires more unitary conjugation but less usage of single mode gates. So as the overall order increases the single mode gates become harder to decompose but the number of operators needed for the unitary conjugation here do not increase as fast. 

A potential problem with this method, as well as the exact decomposition methods of previous sections, are terms with a mix of $\hat{x}$ and $\hat{p}$ operators which act on the same mode. To deal with these terms the commutator expansion method can be used, which was described in section \ref{Commutator}, or one can use another approximate method introduced in the next section.

\begin{table}[!t]
 \centering
 \begin{tabular}{ l|l|l|l } 
 \hline
 Target gate & Commutator approx. & Exact decomposition & Exact decomposition 
 \\ 
 & ($10^{-3}$ precision) & &  
 using Eq.~(\ref{GenProd}, \ref{OddProd})\\
\hline \hline
&&& \\[-7pt]
 $e^{it\hat{x}^4}$ & $1.8\times 10^4$ gates & 29 gates &  - \\ 
\hline
&&& \\[-7pt]
 $e^{it\hat{x}_{j}^2\hat{x}_{k}^2}$ & $2.8\times 10^4$ gates & 119 gates & 279 gates \\ 
 \hline
&&& \\[-7pt]
 $e^{it\hat{x}_{j}\hat{x}_{k}^3}$ & $2.9 \times 10^8$ gates & 269 gates & 93 gates \\ 
 \hline
&&& \\[-7pt]
$e^{it\hat{x}_{j}\hat{x}_{k}\hat{x}_{l}}$ & $4.2\times 10^8$ gates & 17 gates  & 20 gates\\
\hline
&&& \\[-7pt]
$e^{it\hat{x}_{j}^{2}\hat{x}_{k}\hat{x}_{l}}$ & $1.4\times 10^9$ gates & 249 gates  & 198 gates \\
\hline
&&& \\[-7pt]
$e^{it\hat{x}_{j}\hat{x}_{k}\hat{x}_{l}\hat{x}_{m}}$ & $6.9\times 10^{13}$ gates & 440 gates  & 280 gates\\
\hline
&&& \\[-7pt]
$e^{it\hat{x}^6}$ & $1.2\times 10^{13}$ gates & 809 gates &  - \\
\hline
&&& \\[-7pt]
$e^{it\hat{x}_{j}^2\hat{x}_{k}^4}$ & $2.4 \times 10^{13}$ gates & 3320 gates & 12165 gates \\
\hline
\end{tabular}
 \caption{Gate counts for decompositions of some common operations, using the standard commutator approximation as well as the exact decomposition methods demonstrated in this work. The final column uses the generalization for multi-mode exact decompositions using Eq.~(\ref{GenProd}, \ref{OddProd}). The gate counts neglect any Fourier transforms used as they are inexpensive to implement experimentally.}
    \label{ExactVsComm}
\end{table}

\section{Asymptotically exact decompositions with squeezing and displacement} \label{SqDis}

In the last two sections we discussed the exact decompositions of Gaussian operations and a subset of non-Gaussian operations. To be precise, for the latter group we focused on non-Gaussian gates that have generators of monomials that mutually commute.
In this section we discuss some recent ideas on how to decompose more complicated gates in which one mixes in the generators of position and momentum of the same mode. The method was introduced by Yanagimoto et al. \cite{MabuchiKerr}.
Both squeezing and displacement are operations which can be implemented with linear optics. As a result these operations are Gaussian, and are not costly to implement. In fact, the displacement operator is equivalent to a Gaussian operator which is already part of the universal set in Eq.~(\ref{Uniset}). The squeezing operation was given in Eq.~(\ref{sq}), and the displacement operation was given in Eq.~(\ref{eq:Zgatedef}).

In order to use these operations to form a gate decomposition method, we can examine their effect in unitary conjugation with the quadrature operators. Squeezing has the effect of multiplying $\hat{x}$ and $\hat{p}$ operators by a constant
\begin{align}
& \T^{\dagger}(\log\lambda)\hat{x}_{i}\T(\log\lambda) = \lambda \hat{x}_{i}, \\
& \T^{\dagger}(\log\lambda)\hat{p}_{i}\T(\log\lambda) = \lambda^{-1} \hat{p}_{i},
\end{align} 
and displacement has the effect of adding a constant as in Eq.~(\ref{disptransform}) (which will we will specify by $\y$) \cite{Strawberry}. Applying both of these to the annihilation and creation operators maps them to a new effective operation
\begin{equation} 
\hat{a}^{\dagger} \rightarrow \hat{a}^{\dagger}_{\text{eff}} = \T^{\dagger}(\log\lambda)D^{\dagger}(\y)\hat{a}^{\dagger}D(\y)\T(\log\lambda) =  \frac{\lambda}{\sqrt{2\hbar}}\hat{x} - i\frac{\lambda^{-1}}{\sqrt{2\hbar}}\hat{p} + \y.
\end{equation}
With this mapping and the correct choice of constants $\lambda$ and $\y$, terms with a greater overall power in $\hat{x}$ or $\hat{p}$ can be picked out while the contribution of other terms can be ignored. To illustrate this consider transforming the Kerr operation into a cubic phase gate, as in \cite{MabuchiKerr}. Starting with the driven Kerr Hamiltonian
\begin{align}
H_{\text{Kerr}} = -\frac{\chi}{2}\hat{a}^{\dagger^{2}}\hat{a}^{2} + \delta \hat{a}^{\dagger}\hat{a} + \beta (\hat{a} + \hat{a}^{\dagger}),
\end{align}
where $\delta$ and $\beta$ are the detuning and drive constants respectively, and $\chi$ is associated with the self phase modulation of a propagating pulse in a Kerr medium \cite{MabuchiKerr}. We can then apply squeezing and displacement to find the effective Hamiltonian
\begin{align}
H_{\text{Kerr}}^{\text{eff}} =& \T^{\dagger}(\log\lambda)D^{\dagger}(\y) H_{\text{Kerr}} D(\y)\T(\log\lambda) \\
=&-\frac{\chi}{8}(\lambda^{4}\hat{x}^{4} + \hat{p}\hat{x}^{2}\hat{p} + \hat{x}\hat{p}^{2}\hat{x} + \lambda^{-4}\hat{p}^{4}) - \frac{\sqrt{\hbar}}{\sqrt{2}}\chi\lambda^{3}\y\hat{x}^{3} - \frac{\sqrt{\hbar}}{\sqrt{2}}\chi\lambda^{-1}\y\hat{p}\hat{x}\hat{p} \\ \nonumber
&+ \frac{\hbar}{2}\lambda^{2}(-3\chi\y^{2} + \chi + \delta)\hat{x}^{2} + \frac{\hbar}{2}\lambda^{-2}(-\chi\y^{2} + \chi + \delta)\hat{p}^{2} + \sqrt{2}\hbar^{3/2}\lambda(-\chi\y^{3} + \chi\y + \delta\y + \beta)\hat{x}.
\end{align}
Some terms will then cancel by choosing $\delta = 3\chi\y^{2} - \chi$ and $\beta = -2\chi\y^{3}$ to get
\begin{align}
H_{\text{Kerr}} = -\frac{\chi\lambda^{3}\y}{\sqrt{2}}\left(\hat{x}^{3} + \frac{\lambda\y^{-1}}{4\sqrt{2}}\hat{x}^{4} - \sqrt{2\hbar}\lambda^{-5}\y \hat{p}^{2} + \mathcal{O}(\lambda^{-4}) \right).
\end{align}
With the correct choice of the displacement constant $y$, namely $y \sim \lambda^{3}$, the effective Hamiltonian takes on the form of a cubic operation to leading order in $\lambda$. Although this decomposition can be done with only squeezing and displacement, and a tuning of their respective parameters, we note that squeezing to arbitrary levels can be difficult to implement \cite{braunstein2005squeezing}. \\

\section{Discussion} \label{discussion}

In this work we outlined approximate and exact gate decomposition methods for CV quantum computers. For approximate methods we reviewed a commutator expansion method which expressed quadrature operators in terms of commutators, as well as another decomposition procedure which used squeezing and displacement operations in unitary conjugation to approximate gates. We also examined exact decomposition methods both for Gaussian and non-Gaussian operations. For Gaussian operators we showed how to express the target gate in terms of displacement, interferometer and squeezing gates. In order to decompose non-Gaussian operations exactly we used a technique where the target gate was expressed in terms of a linear combination of polynomials and then decomposed using unitary conjugation, a lemma to the Baker-Campbell-Hausdorff formula (Eq.~(\ref{Eq:Unit_conj}) and (\ref{Eq:BCH})) and other known lower order exact decompositions. This procedure was generalized using Eq.~(\ref{GenProd}, \ref{OddProd}) to allow for any product of quadrature operators to be expressed as a linear combination of polynomials of the same operators. To illustrate these methods we examined examples of applying them to some simpler operations. The decompositions were all made given the chosen universal set in Eq.~(\ref{Uniset}).

The commutator expansion method provided a comprehensive CV decomposition method that allows for the decomposition of any operator into gates from a universal set. Although it allowed for the decomposition of any operator, the recursive structure of the method causes exponential build up in the number of gates required as the order of the operator to be decomposed increases. Combined with the need to repeat the final sequence of gates to increase the precision of the decomposition, the overall circuit depth may become very large. As an example we examined the decomposition of the gate $e^{it\left(\hat{x}^2\hat{p} + \hat{p}\hat{x}^2\right)}$ in Eq.~(\ref{commexample}), as well as the Kerr operation. For these examples the method provided decompositions requiring small sequences of gates from the universal set, but in order to achieve a modest precision, the product of gates needed to be repeated many times.

While more limited in the scope of what can be decomposed for non-Gaussian gates, exact decompositions can in some cases significantly lower the gate count needed for a decomposition when compared to approximate methods. For example, the gate used as an example of applying the commutator expansion method ($e^{it\left(\hat{x}^2\hat{p} + \hat{p}\hat{x}^2\right)}$) cannot be decomposed exactly with the methods shown in this work, due to the mixture of $\hat{x}$ and $\hat{p}$ quadrature operators of the same mode. Although another multi-mode operation, namely $e^{it\hat{x}_{j}\hat{x}_{k}\hat{x}_{l}}$, can be decomposed exactly with only 17 gates from the universal set, whereas it would take $4.2\times 10^8$ gates using the commutator expansion method for a precision of $10^{-3}$. The set of gates for which exact decompositions can be obtained still encompass a large class of useful CV operators \cite{sparrow2018simulating, CVHHL, kalajdzievski2018continuous, BHExtended, CommutatorApprox, ChrisAlgo2017, MonteCarlo}, and can be expanded by using Eq.~(\ref{GenProd}, \ref{OddProd}). This expanded method for multi-mode operations can also be used to decompose operations that could already be decomposed with the regular exact method to compare gate counts. Some example gate counts can be seen in Table \ref{ExactVsComm}. In some cases it seems as though the expanded method requires more unitary conjugation but less usage of single-mode gates and as a result can require fewer gates. It is still unclear whether exact decompositions can be generalized to include all CV operations.

The final method examined in this work uses squeezing and displacement operators in unitary conjugation to approximate an operator into another contained within the universal set. While no longer being exact, this method allowed for the precision of the decomposition to be increased by simply tuning the parameters of the final sequence of gates instead of having to repeat them. To demonstrate this method we examined how the Kerr operator can be approximated in terms of a cubic gate, following the procedure in \cite{MabuchiKerr}.  

By putting together all these results in a single place, it is our hope that this manuscript becomes an easy to use reference for researchers in the area, and moreover, that it inspires further work in finding more efficient and compact decompositions of CV gates.

\section*{Acknowledgments}
T.K. thanks C. Weedbrook for valuable discussions. T.K. and N.Q. thank N. Killoran and L. Madsen for comments on the manuscript.

\bibliographystyle{unsrtnat}
\bibliography{DS}

\end{document}